\documentclass[aps,floatfix,amsmath,amssymb,nofootinbib,twocolumn]{revtex4-1}
\usepackage{graphicx}
\usepackage{dcolumn}
\usepackage{bbm}
\usepackage{epstopdf}
\usepackage[bookmarks=true,colorlinks,linkcolor=blue,urlcolor=blue,citecolor=blue]{hyperref}
\usepackage{bbold}
\usepackage{color}
\usepackage[toc,page]{appendix}
\usepackage[percent]{overpic}
\usepackage{enumitem}
\usepackage{graphicx}
\usepackage{tikz}
\usepackage{subcaption}
\usepackage{setspace} 
\newcommand{\beq}{\begin{equation}}
\newcommand{\eeq}{\end{equation}}
\newcommand{\bal}{\begin{align}}
\newcommand{\eal}{\end{align}}
\newcommand{\beqn}{\begin{eqnarray}}
\newcommand{\eeqn}{\end{eqnarray}}

\usepackage{amsmath}
\usepackage{amssymb}
\newcommand{\RNum}[1]{\uppercase\expandafter{\romannumeral #1\relax}}

\begin{document}

\title{A slave boson description of pseudogap metals in $t$-$J$ models}

\author{Julia Brunkert}
\author{Matthias Punk}
\affiliation{Physics Department, Arnold Sommerfeld Center for Theoretical Physics, Center for NanoScience, and Munich Center for Quantum Science and Technology (MCQST), Ludwig-Maximilians University Munich, Germany}

\date{\today}

\begin{abstract}
We present a simple modification of the standard $U(1)$ slave boson construction for the single band $t$-$J$ model which accounts for two-particle bound states of spinons and holons. This construction naturally gives rise to fractionalized Fermi liquid ground states, featuring small, hole-like pocket Fermi surfaces with an anisotropic quasiparticle weight in the absence of broken symmetries. 
In a specific parameter regime our approach maps the square lattice $t$-$J$ model to a generalized quantum dimer model, which was introduced as a toy model for the metallic pseudogap phase in hole-doped cuprates in [PNAS 112,9552-9557 (2015)]. Our slave boson construction captures essential features of the nodal-antinodal dichotomy and straightforwardly describes sharp, Fermi arc-like features in the electron spectral function. Moreover, it allows to study quantum phase transitions between fractionalized Fermi liquid phases and superconductors or ordinary Fermi liquids.
\end{abstract}

\maketitle 

 \section{Introduction}
Even though the single band Hubbard model and its strong coupling descendant, the  $t$-$J$ model, are among the most basic lattice models for interacting electrons, relatively little is known about their ground state properties at electron densities slightly away from the Mott-insulator at half filling. In the strongly interacting regime an intricate interplay between spin and charge degrees of freedom can give rise to a plethora of different possible ground states, including various symmetry broken states with or without spin and/or charge order, depending on the lattice geometry as well microscopic details of the electron hopping parameters. The relevance of these models for the description of real materials mainly derives from the cuprate high-temperature superconductors, where the $t$-$J$ model on the square lattice is believed to capture essential correlation properties of electrons in the CuO$_2$ planes \cite{Dagotto1994, Ogata2008}.

While large scale numerical simulations for these models are challenging due to the fermion sign problem, progress has been made in particular using dynamical mean-field theory and its cluster extensions \cite{Maier2005, Kotliar2006}, as well as diagrammatic Monte-Carlo methods \cite{Prokofev2007}. Applied to the single-band square-lattice Hubbard model in two-dimensions, both methods showed that a pseudogap develops below half filling for sufficiently strong interactions, i.e.~the electronic density of states is suppressed in the antinodal regions of the Brillouin zone close to $\mathbf{k}=(0,\pi)/a$ and symmetry related momenta \cite{Macridin2006, Civelli2008, Sordi2012, Gull2013, Wu2017}.

In order to study some of the proposed phases and their properties, the slave boson approach has proven to be a very valuable analytical tool \cite{Barnes1976, Kotliar1986, Kotliar1988, Suzumura1988, Li1989, LeeNagaosaWen}. In this parton construction the electron operator is represented in terms of a fermionic operator carrying the electron spin, as well as a bosonic operator which keeps track of the missing electron charge relative to the half filled case. The $t$-$J$ model then takes the form of a gauge theory describing fermionic spinon as well as bosonic holon degrees of freedom and their mutual, gauge field mediated interaction \cite{LeeNagaosaWen}.
One problem with this approach is that experimental signs of spin-charge separation in the underdoped cuprates are inconclusive. Transport and spectroscopic experiments suggest that at least some of the low energy excitations in the metallic pseudogap phase are electron-like, carrying both spin and charge \cite{Mirzaei2013, Chan2014, Yoshida2012}. In particular, within a simple mean-field picture of spin-charge separation, the electron spectral function is a convolution of the spinon and holon spectral functions, which cannot give rise to the sharp Fermi arcs observed in angle-resolved photo emission experiments (ARPES) \cite{Norman1998, Shen2005, Yang2011}. 

A possible solution to this problem is that spinons and holons form two-particle bound states which carry both spin and charge \cite{Beran1996}. While gauge field fluctuations do mediate an attractive interaction, simple electron hopping can lead to bound state formation as well. Different approaches have been developed to account for spinon-holon bound states in parton constructions for the $t$-$J$ model, such as Ribeiro and Wen's spinon-dopon approach \cite{Ribeiro2005, Ribeiro2006}, or the phenomenological description of such bound states by Ng \cite{Ng2005}. While the former introduces new auxiliary degrees of freedom leading to a more complex representation of the electron operator, the latter studies consequences of a phenomenological attractive spinon-holon interaction within the standard U(1) slave boson framework. In both cases the results are in better agreement with experimental observations.

In this work we show that spinon-holon bound states can be straightforwardly incorporated in the standard $U(1)$ slave particle approach. The main difference to Ribeiro and Wen's approach is that these bound states live on the links between two lattice sites, rather than individual lattice sites. This has important consequences for the electronic quasiparticle weight and is the reason for the appearance of Fermi arc-like features in the electron spectral function, as we discuss in detail below. In the absence of broken symmetries one ground state of our model is a fractionalized Fermi liquid (FL*) \cite{SenthilVojtaSachdev}, where these hole-like bound states form a small Fermi surface. This exotic metallic phase obeys a modified Luttinger count with a Fermi volume proportional to the density of doped holes $p$ away from the Mott insulator at half filling \cite{SenthilVojtaSachdev2, Paramekanti2004, Sachdev2016, Sachdev2018}, rather than the total density of holes $1+p$ measured from the filled band as in ordinary Fermi liquids. Such a small Fermi volume is consistent with the Drude spectral weight and Hall resistivity measurements in the pseudogap phase of the cuprates \cite{Orenstein1990, Uchida1991, Ando2004, Balakirev2009, Badoux2015}.

FL* ground states of the $t$-$J$ model have been discussed previously using the spinon-dopon approach \cite{Mei2012, Punk2012} and in terms of a generalized quantum dimer model \cite{Punk2015, Goldstein2017, Huber2018, Feldmeier2018}. While the former does give rise to a Fermi surface with small hole pockets close to momenta $\mathbf{k} = (\pm \pi/2, \pm \pi/2)/a$, the electronic quasiparticle weight is relatively uniform around the Fermi surface. By contrast, the dimer model has spinon-holon bound states living on nearest neighbor links. This leads to similar hole pockets as the spinon-dopon approach, but with a strongly anisotropic quasiparticle weight around the Fermi pockets, which would appear as Fermi arcs in photoemission experiments.

In this work we show that our modified slave boson construction maps to the above mentioned dimer model in a specific parameter regime, where the ground state is a $U(1)$-FL* with a propagating, emergent photon-like mode. Moreover, our approach allows for an exceptionally simple description of $\mathbb{Z}_2$-FL* phases, where the $U(1)$ gauge field is gapped due to the presence of a spinon pair condensate. This phase features an electron spectral function with the same qualitative features as the $U(1)$-FL*. The main advantage of our slave boson construction compared to the above mentioned previous approaches is that it allows to study quantum phase transitions between the FL* pseudogap phases and a superconductor or an ordinary Fermi liquid. We mention here that a different parton construction has been developed recently, where this is possible as well \cite{Zhang2020}. 

The remaining article is structured as follows: in Sec.~\ref{sec:intro} we present the simple modification of the $U(1)$ slave boson approach which accounts for spinon-holon bound states. Starting from this theory we derive an effective model for $U(1)$- and $\mathbb{Z}_2$-FL* ground states in Sec.~\ref{sec:fls} and discuss their properties within a simple saddle-point approximation. In Sec.~\ref{sec:fluct} we study important gauge fluctuations and derive low energy theories for these phases. Finally, in Sec.~\ref{sec:trans} we propose theories for the quantum phase transition between the $\mathbb{Z}_2$-FL* and a superconductor, as well as for the transition from an $U(1)$-FL* to an ordinary Fermi liquid and point out potential problems with the latter.

\section{U(1) slave boson construction}
\label{sec:intro}

Our starting point is the Hamiltonian of the $t$-$J$ model 
\begin{equation}
H = - \sum_{i,j,\sigma} t_{ij} c^\dagger_{i \sigma} c^{\ }_{j \sigma} + J \sum_{\langle i,j \rangle} \left( \mathbf{S_i} \cdot \mathbf{S_j} - \frac{1}{4} n_i n_j \right) \ ,
\label{tjmodel}
\end{equation}
where $c_{j \sigma}$ is a Gutzwiller projected electron operator on lattice site $j$ (i.e.~doubly occupied sites are projected out), $\sigma=\uparrow,\downarrow$ denotes the electron spin, $t_{ij}$ are the electron hopping amplitudes, $\mathbf{S}_j$ is the electron spin operator and $n_j$ the density of electrons on lattice site $j$.

In the standard U(1) slave boson construction the Gutzwiller projected electron creation operator is represented as \cite{LeeNagaosaWen}
\begin{equation}
c^\dagger_{i \sigma} = f^\dagger_{i \sigma} b_i \ .
\label{eq1}
\end{equation}
Here $f^\dagger_{i \sigma}$ is a fermionic spinon creation operator, whereas the bosonic operator $b_i$ destroys a holon on lattice site $i$ and accounts for missing charge below half filling. In order for Eq.~\eqref{eq1} to hold, the particle number constraint
\begin{equation}
n^f_{i \uparrow}+n^f_{i \downarrow} + n^b_i =1
\label{constraint}
\end{equation}
has to be imposed on each lattice site $i$, where $n^f$ and $n^b$ denote the fermion and boson density operators, respectively. Also note that the slave boson representation in  has a local $U(1)$ gauge redundancy and the electron creation operator in Eq.~\eqref{eq1} is invariant under the gauge transformation
\begin{equation}
f_j \to f_j e^{i \phi_j} \ , \hspace{1cm} b_j \to b_j e^{i \phi_j} \ ,
\label{gauge}
\end{equation}
where $\phi_j$ is an arbitrary, lattice site dependent phase.
After decoupling the Heisenberg interaction term in the hopping and pairing channel, the Lagrangian of the $t$-$J$ model in imaginary time $\tau$ takes the form (see e.g.~Ref.~\cite{LeeNagaosaWen}),
\begin{eqnarray}
\mathcal{L} &=& \sum_{i,\sigma} \bar{f}_{i \sigma} \left( \partial_\tau-i \lambda_i \right) f_{i \sigma} + \sum_i \bar{b}_{i} \left( \partial_\tau-i \lambda_i +\mu_B \right) b_{i} \notag \\
&&- \tilde{J} \sum_{\langle i,j \rangle} \left[ \bar{\chi}_{ij} \bar{f}_{i \sigma} f_{j \sigma} + h.c. - |\chi_{ij}|^2 \right] \notag \\
&&+ \tilde{J}  \sum_{\langle i,j \rangle} \left[ \bar{\Delta}_{ij} (f_{i \uparrow} f_{j \downarrow} - f_{i \downarrow} f_{j \uparrow})  + h.c. + |\Delta_{ij}|^2 \right] \notag \\
&& - \sum_{i,j} t_{ij} \, \bar{f}_{i \sigma} b_i \bar{b}_j f_{j \sigma} \ ,
\label{lagrangian1}
\end{eqnarray}
where $\lambda_i$ is a Lagrange multiplier that enforces the constraint Eq.~\eqref{constraint} and $\chi_{ij}$ ($\Delta_{ij}$) are spinon hopping (pairing) bond fields which have been used to decouple the four fermion spin-spin interaction term \cite{Baskaran1987}. The overbar denotes complex conjugation for bosonic fields and $\tilde{J} = 3 J/8$ is a renormalized exchange coupling \cite{Brinckmann2001}. Moreover, the $\sim n_i n_j$ in Eq.~\eqref{tjmodel}, which can be written as a nearest neighbor holon-holon interaction, was neglected as usual, because it is not expected to play an important role at small hole doping. Usually the term in the last line of Eq.~\eqref{lagrangian1}, which derives from the electron hopping term in Eq.~\eqref{tjmodel}, is decoupled using the spinon hopping field $\chi_{ij}$ as well. The resulting theory is a common starting point for the construction of mean field phase diagrams and different phases can be straightforwardly obtained by condensing combinations of the bosonic fields $b$, $\Delta$ and $\chi$. The pseudogap phase in underdoped cuprates is then identified with the phase where $\langle \chi \rangle \neq 0$ and $\langle \Delta \rangle \neq 0$, but the holons are not condensed ($ \langle b \rangle = 0$) \cite{Fukuyama1992, LeeNagaosaWen}.

As mentioned in the introduction, a major problem with this description of the pseudogap phase is that deconfined spinons and holons are the low-energy degrees of freedom, whereas transport and spectroscopic measurements in underdoped cuprates indicate that some excitations are electron- or hole-like. For this reason we want to introduce hole-like bound states of spinons and holons in Eq.~\eqref{lagrangian1}, which carry both electric charge and spin. It is important to emphasize here that the attraction between spinons and holons which gives rise to this bound state is assumed to be due to electron hopping and the theory remains deconfined, i.e.~spinon excitations are still allowed to propagate.

The main idea of our work is that such bound states can be introduced straightforwardly via a decoupling of the electron hopping term in the last line of Eq.~\eqref{lagrangian1} using \emph{fermionic} Hubbard-Stratonovich bond fields $F_{ij\sigma}$, $\bar{F}_{ij\sigma}$ which carry both spin and electric charge, as well as a $U(1)$ gauge charge of two: 
\begin{equation}
\bar{F}_{ij\sigma} \equiv \left( \bar{f}_{i \sigma} \bar{b}_j + \bar{f}_{j \sigma} \bar{b}_i \right)/\sqrt{2} \ .
\end{equation}
The field $F_{ij\sigma}$ represents such a fermionic hole-like bound state and naturally lives on the lattice bonds between sites $i$ and $j$. Note that fermionic excitations carrying both spin and electric charge cannot exist on a single lattice site due to the constraint in Eq.~\eqref{constraint}. 
After decoupling the electron hopping term, the Lagrangian Eq.~\eqref{lagrangian1} takes the form
\begin{eqnarray}
\mathcal{L} &=& \sum_{i,\sigma} \bar{f}_{i \sigma} \left( \partial_\tau-i \lambda_i \right) f_{i \sigma} + \sum_i \bar{b}_{i} \left( \partial_\tau-i \lambda_i +\mu_B \right) b_{i} \notag \\
&&- \tilde{J} \sum_{\langle i,j \rangle} \left[ \bar{\chi}_{ij} \bar{f}_{i \sigma} f_{j \sigma} + h.c. - |\chi_{ij}|^2 \right] \notag \\
&&+ \tilde{J}  \sum_{\langle i,j \rangle} \left[ \bar{\Delta}_{ij} (f_{i \uparrow} f_{j \downarrow} - f_{i \downarrow} f_{j \uparrow})  + h.c. + |\Delta_{ij}|^2 \right] \notag \\
&& + \sum_{i,j} t_{ij} \, \bar{F}_{ij\sigma} F_{ij \sigma} \notag \\
&& + \sum_{i,j} \frac{t_{ij}}{\sqrt{2}}  \left[ \bar{F}_{ij\sigma} \left( f_{i \sigma} b_j + f_{j \sigma} b_i \right) + h.c. \right] \ .
\label{lagrangian2}
\end{eqnarray}
Note that upon integrating out the fermions $F_{ij\sigma}$ one recovers Eq.~\eqref{lagrangian1} with an additional interaction term $\sim n^f_i n^b_j$, which can be expressed as a holon-holon interaction using the constraint Eq.~\eqref{constraint}. This interaction will be neglected in the following, in analogy to the $\sim n_i n_j$ term in the derivation of Eq.~\eqref{lagrangian1} from the $t-J$ model. 

\begin{figure}
\includegraphics[width=0.8 \columnwidth]{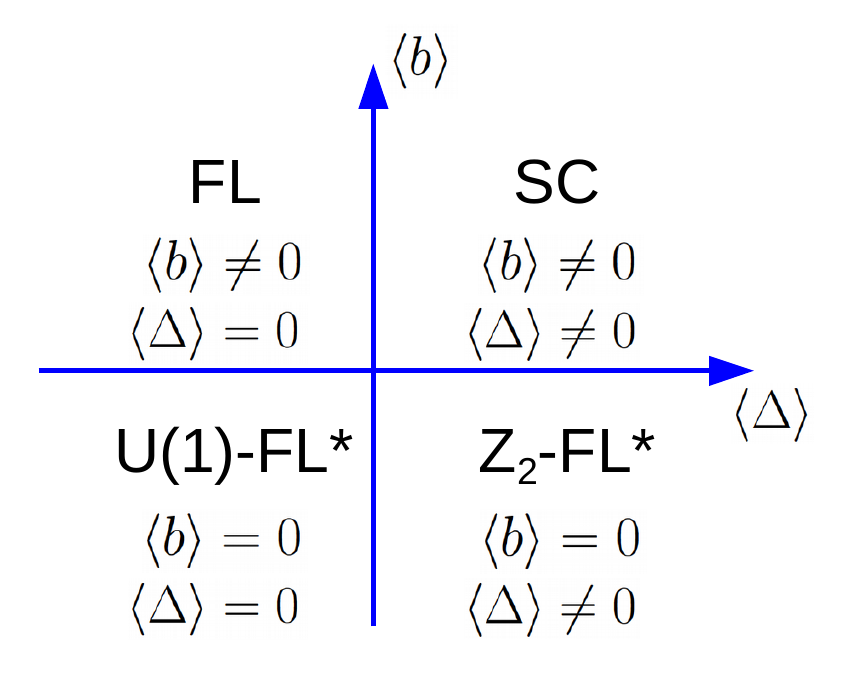}
\caption{Schematic phase diagram of the model in Eq.~\eqref{lagrangian2} at a finite holon density and for $\langle \chi \rangle \neq 0$. FL and SC denote ordinary Fermi liquid and superconducting phases, respectively, whereas both phases with $\langle b \rangle = 0$ correspond to fractionalized Fermi liquids (FL*), discussed in detail in Sec.~\ref{sec:fls}.}
\label{fig0}
\end{figure}

The model in Eq.~\eqref{lagrangian2} has a very similar structure as the theory for Kondo-Heisenberg models studied in Ref.~\cite{SenthilVojtaSachdev2}, where the conduction electrons take the role of our $F$ fermions. The main difference is that our $F$ fermions carry a $U(1)$ gauge charge of two, whereas the conduction electrons in Ref.~\cite{SenthilVojtaSachdev2} are gauge neutral. Nevertheless, in our case the Fermi surface of $F$ fermions coincides with the physical electron Fermi surface in the FL* phases, despite the fact that our $F$ fermions are gauge charged. 

In the following we are interested in symmetric phases of Eq.~\eqref{lagrangian2}, in particular in the regime where $\langle \chi \rangle  \neq 0$, corresponding to resonating valence bond (RVB) states in the undoped (i.e.~half filled) case \cite{Baskaran1987}. The structure of the phase diagram is then determined by the holon condensate $\langle b \rangle$ as well as the spinon pair condensate $\langle \Delta \rangle$ and is sketched in Fig.~\ref{fig0}. If the holons are condensed, $\langle b \rangle \neq 0$, the situation is analogous to the standard $U(1)$ slave boson construction: we have $c^\dagger_{i\sigma} \sim f^\dagger_{i\sigma}$ from Eq.~\eqref{eq1} and the ground state is either an ordinary Fermi liquid for $\langle \Delta \rangle = 0$, or a superconductor for $\langle \Delta \rangle \neq 0$. For this reason we focus on interesting fractionalized phases with $\langle b \rangle = 0$ in the following, where we get different ground states compared to the standard slave boson approach.

\section{Doped RVB phases: $U(1)$ and $\mathbb{Z}_2$-FL*}
\label{sec:fls}

The Lagrangian in Eq.~\eqref{lagrangian2} allows for a simple description of doped RVB phases with well defined electronic quasiparticle excitations. These phases appear for $\langle b \rangle =0$ and are distinguished by the presence or absence of a spinon pair-condensate $\langle \Delta \rangle $. As we argue below, both phases are fractionalized Fermi liquids (FL*) with small pocket Fermi surfaces and an anisotropic quasiparticle weight in the absence of broken symmetries. 
These states differ in the nature of their gauge excitations, discussed in Sec.~\ref{sec:fluct}. Whereas the $U(1)$-FL* phase features photon-like degrees of freedom, a non-zero spinon pair condensate $\Delta$ gaps out this photon mode via the Higgs mechanism and we obtain a $\mathbb{Z}_2$-FL* phase. In the following we derive effective theories for both phases and discuss them in detail.

For $\langle b_i \rangle = 0$ the $f$ fermions and the $b$ bosons in Eq.~\eqref{lagrangian2} can be integrated out.
The resulting effective action for  the bond fields $\chi_{ij}$,  $\Delta_{ij}$ and $F_{ij \sigma}$ is strongly constrained by electric charge conservation as well as invariance under the gauge transformation in Eq.~\eqref{gauge}, under which these fields transform as 
\begin{eqnarray}
\Delta_{ij} &\to& \Delta_{ij} \, e^{i (\phi_i+\phi_j)} \\ 
F_{ij\sigma} &\to& F_{ij\sigma} \, e^{i (\phi_i+\phi_j)} \\
\chi_{ij} &\to& \chi_{ij} \, e^{i (-\phi_i+\phi_j)} \ .
\end{eqnarray}
Consequently both, the spinon pairing field $\Delta_{ij}$ as well as the fermionic field $F_{ij\sigma}$ carry a gauge charge of two, while $\chi_{ij}$ has no net gauge charge. The effective Lagrangian takes the form
\begin{widetext}
\begin{eqnarray}
\mathcal{L}_\text{eff} [\chi_{ij}, \Delta_{ij}, F_{ij\sigma}]
&=& \sum_{i,j} \big\{  \bar{\chi}_{ij} \left[ \partial_\tau - i (-\lambda_i+\lambda_j) \right]  \chi_{ij} + a^\chi_{1} |\chi_{ij}|^2 + a^\chi_{2} |\chi_{ij}|^4 \big\} + a^\chi_{3} \sum_{i,j,k,l} \chi_{ij}\chi_{jk} \chi_{kl} \chi_{li} \notag \\
& \ \ +& \sum_{i,j} \big\{  \bar{\Delta}_{ij} \left[ \partial_\tau - i (\lambda_i+\lambda_j) \right]  \Delta_{ij} + a^\Delta_{1} |\Delta_{ij}|^2 + a^\Delta_{2} |\Delta_{ij}|^4 \big\} + a^\Delta_{3} \sum_{i,j,k,l}\bar{\Delta}_{ij} \Delta_{jk} \bar{\Delta}_{kl} \Delta_{li} \notag \\
& \ \ +&  \sum_{i,j} \big\{ \bar{F}_{ij \sigma} \left[ \partial_\tau - i (\lambda_i+\lambda_j) \right] F_{ij \sigma} +a^F_{1} \bar{F}_{ij \sigma} F_{ij \sigma} \big\} \notag \\
& \ \ +&  \sum_{i,j,k,l} \big\{ a^{F\chi}_1 \, \bar{F}_{ij\sigma} F_{jk\sigma} \chi_{kl} \chi_{li} + a^{F\chi}_2 \, \bar{F}_{ij\sigma} \bar{\chi}_{jk} F_{kl\sigma}  \chi_{li} + a^{\Delta \chi}_1 \, \bar{\Delta}_{ij} \Delta_{jk} \chi_{kl} \chi_{li} + a^{\Delta \chi}_2 \, \bar{\Delta}_{ij} \bar{\chi}_{jk} \Delta_{kl}  \chi_{li} \big\} \notag \\
& \ \ +&  \sum_{i,j,k,l} \big\{ a^{F \Delta}_1 \,  \bar{F}_{ij\sigma} F_{jk\sigma} \bar{\Delta}_{kl} \Delta_{li} + a^{F \Delta}_2 \,  \bar{F}_{ij\sigma} \Delta_{ij} \bar{\Delta}_{kl} F_{kl\sigma} \big\} + \dots \ ,
\label{FLstaraction}
\end{eqnarray}
\end{widetext}
where $a^{\cdot}_i$ are real coefficients and a summation over repeated spin indices $\sigma$ is implied. The interesting terms in this Lagrangian involve products of the fields around closed loops and the dots represent allowed higher order terms, in particular products of fields around larger loops which obey charge conservation. Note that loop terms involving only $\Delta_{ij}$ and $F_{ij\sigma}$ always have to contain an even number of fields due to gauge invariance. By contrast, loop terms involving the field $\chi_{ij}$ may depend on an odd number of fields as well, but we are going to limit our remaining discussion to the square lattice case with the fields $\chi_{ij}$, $\Delta_{ij}$ and $F_{ij\sigma}$ restricted to nearest neighbor bonds, where odd terms cannot appear. In this case all explicitly shown loop terms in Eq.~\eqref{FLstaraction} are defined on elementary plaquettes of the square lattice. Also note that the density of $F$ fermions is fixed by the density of holons, i.e.~by the density of holes away from half filling.

The action in Eq.~\eqref{FLstaraction} admits several saddle point solutions and in the following we are only interested in translationally and rotationally invariant phases with $\langle \chi \rangle \neq 0$ (i.e.~RVB phases). For the square lattice case several different saddle points for $\chi_{ij}$ have been discussed in the literature and we limit our analysis in this work to the simple uniform RVB state where $\langle \chi_{ij} \rangle = \chi \in \mathbb{R}$ on all nearest neighbor bonds. We note here that we do not expect qualitative differences in electronic properties for the widely-discussed $\pi$-flux state \cite{Affleck1988, Dagotto1988}, where $\langle \chi_{ij} \rangle = \chi \exp[ i (-1)^{i_x+j_y} \pi/4]$. This is because, as shown below, electronic observables are tied to properties of the $F$ fermions, which carry a gauge charge of two. In the $\pi$-flux phase the $F$ fermions thus pick up an Aharonov-Bohm phase of $2 \pi$ when encircling an elementary plaquette, which leaves their low energy properties unchanged. Fluctuations of $\chi_{ij}$ beyond the mean-field solution are discussed in Sec.~\ref{sec:fluct}.

\begin{center}
\begin{figure}
\includegraphics[width=0.7 \columnwidth]{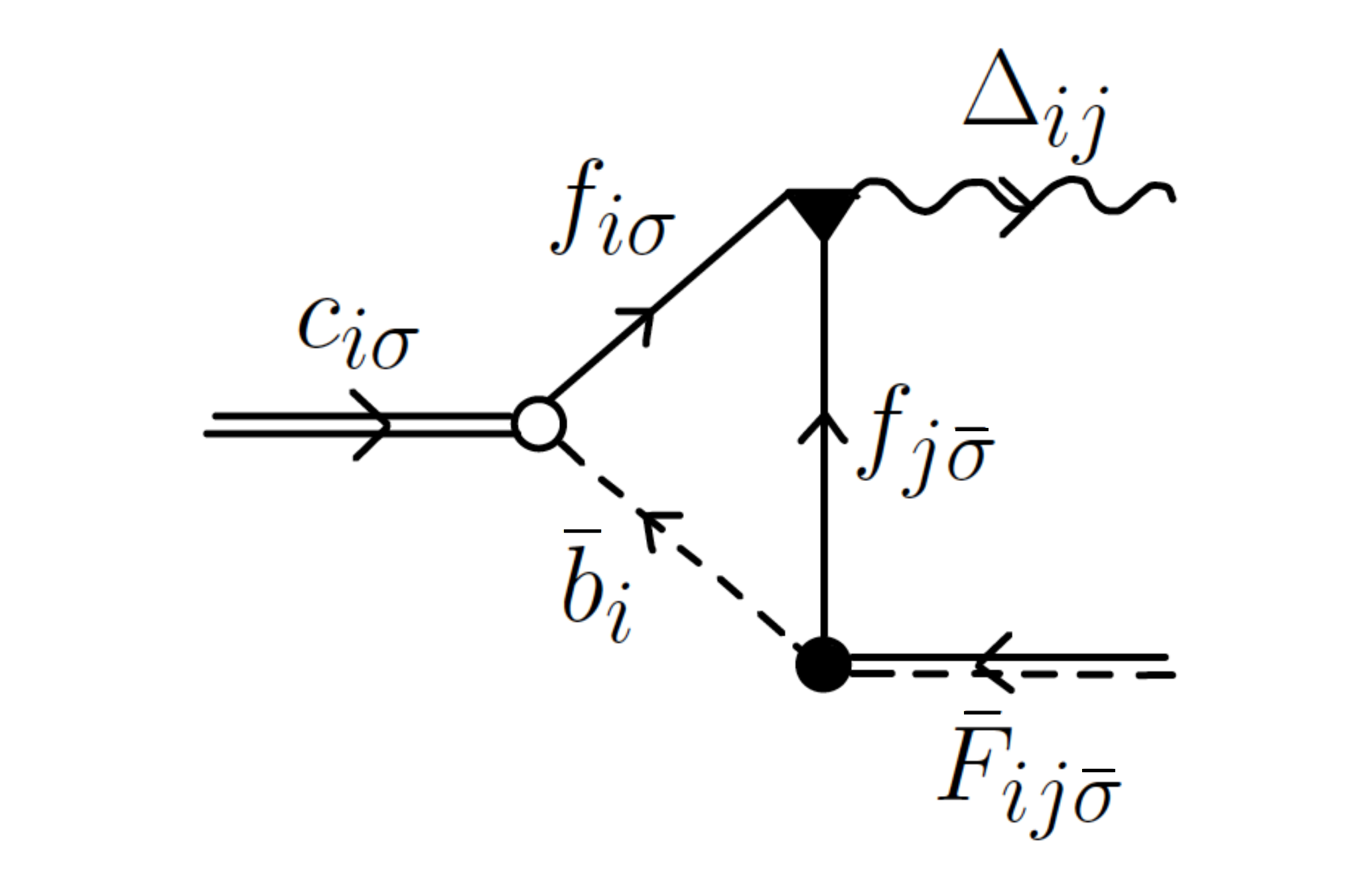}
\caption{Electron field $c_{i\sigma}$ in terms of $\Delta_{ij}$ and $\bar{F}_{ij \bar{\sigma}}$. }
\label{fig1}
\end{figure}
\end{center}

Ultimately we are interested in electronic properties of the model in Eq.~\eqref{FLstaraction}. In the saddle point approximation for $\chi_{ij}$ the gauge invariant electron field $c_{i\sigma}$ can be uniquely expressed in terms of the bond fields $\Delta_{ij}$ and $F_{ij\sigma}$ as (see Fig.~\ref{fig1})
\begin{equation}
c_{i \sigma} \sim  \sum_j \bar{F}_{i j \bar{\sigma}} \Delta_{ij} \ , 
\label{electron}
\end{equation}
where $\bar{\sigma}$ denotes the opposite spin of $\sigma$. This important relation will be used later to compute electron spectral functions.

Within the manifold of saddle points with fixed $\chi$, two simple symmetric phases can be realized in the model Eq.~\eqref{FLstaraction} on the square lattice. For $\langle \Delta_{ij} \rangle \neq 0$ we obtain a $\mathbb{Z}_2$-FL* phase. In this case Eq.~\eqref{electron} implies that $c_{i \sigma} \sim  \sum_j \bar{F}_{i j \bar{\sigma}}$ and thus the electronic Fermi surface coincides with the small Fermi surface of $F$ fermions. This implies a modified Luttinger count of the Fermi volume, which is proportional to the density of $F$ fermions, i.e.~the density of doped holes away from half filling.
On the other hand, for  $\langle \Delta_{ij} \rangle = 0$ we realize a $U(1)$-FL*, which also features a sharp electronic Fermi surface, despite the fact that Eq.~\eqref{electron} seemingly implies that the electron spectral function is a convolution of the $F_{ij\sigma}$ and $\Delta_{ij}$ spectral functions. This state also features a small Fermi surface.

\subsection{$U(1)$-FL* on the square lattice}
\label{sec:u1fls}

\begin{center}
\begin{figure}
\includegraphics[width=0.7 \columnwidth]{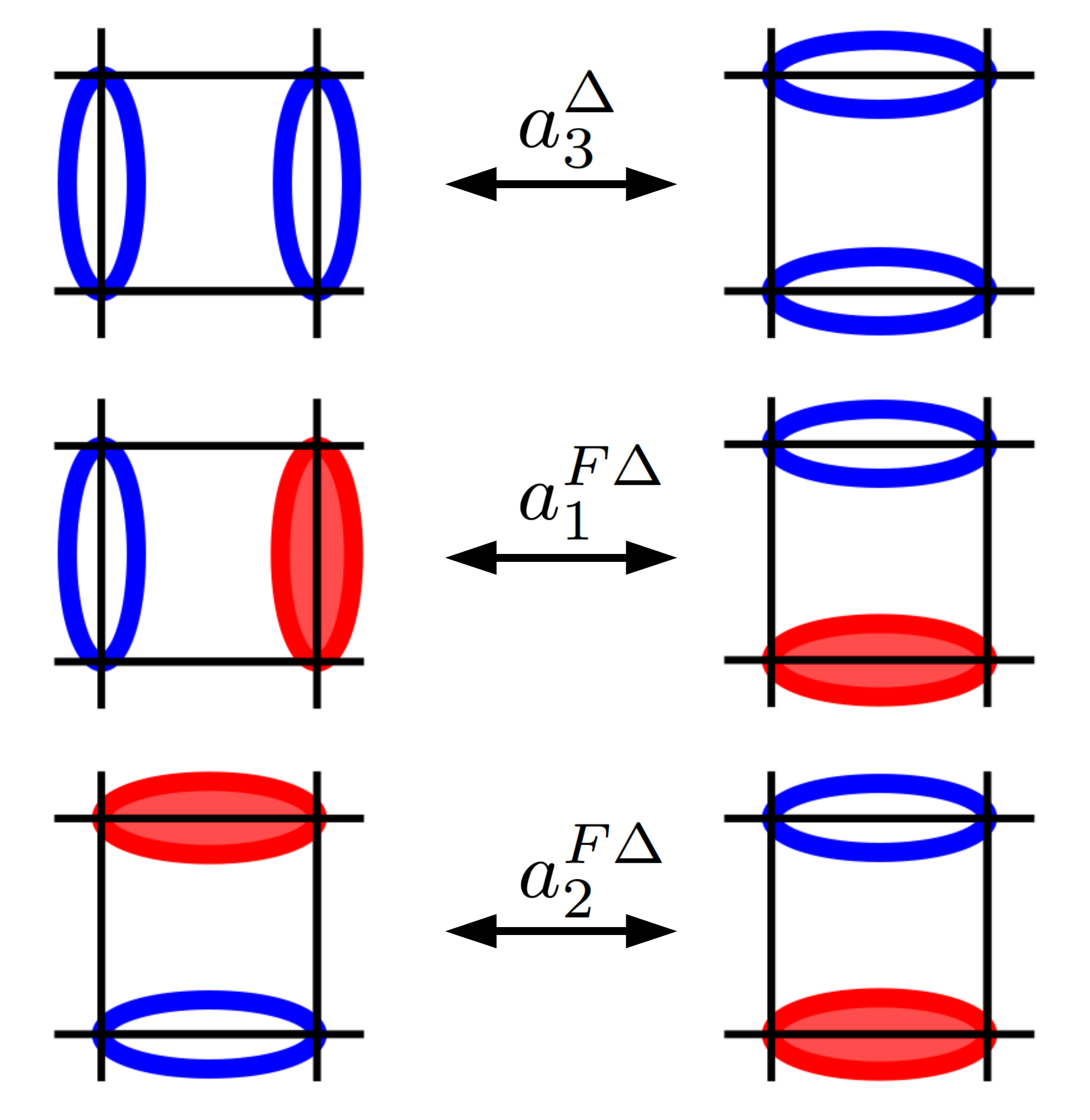}
\caption{Graphic illustration of the dimer resonance terms on an elementary square plaquette in the effective model for the $U(1)$-FL* phase, Eq.~\eqref{dimer}. Empty blue ellipses denote spin-singlets, full red ellipses represent spinon-holon bound states with charge $q=+e$ and spin-$\tfrac{1}{2}$. Symmetry related terms are not shown.}
\label{fig:dimer}
\end{figure}
\end{center}

Let us study the theory in Eq.~\eqref{FLstaraction} on the square lattice at the above mentioned uniform RVB saddle point $\langle \chi_{ij} \rangle =\chi$. We restrict the fields $\Delta_{ij}$ and $F_{ij\sigma}$ to nearest neighbor bonds and set the lattice constant to unity throughout the rest of this work. On the square lattice it is convenient to re-label the fields as $\Delta_{j,\eta}$ and $F_{j,\eta,\sigma}$, where the index $\eta \in \{x,y\}$ determines if the field lives on the bond emanating in positive $x$ or $y$ direction from lattice site $j$. In this case the important loop terms from Eq.~\eqref{FLstaraction} take the form of an interaction Hamiltonian
\begin{eqnarray}
 && a^\Delta_{3} \sum_{j} \Delta^\dagger_{j,x}  \Delta^\dagger_{j+\hat{y},x}  \Delta^{\ }_{j+\hat{x},y} \Delta^{\ }_{j,y} + h.c. \notag \\
+ && a^{F \Delta}_1 \sum_{j,\sigma} F^\dagger_{j,x,\sigma}  \Delta^\dagger_{j+\hat{y},x} \Delta^{\ }_{j,y} F^{\ }_{j+\hat{x},y,\sigma}+ \dots \notag \\
+ && a^{F \Delta}_2 \sum_{j,\sigma} F^\dagger_{j,x,\sigma}  \Delta^\dagger_{j+\hat{y},x} \Delta^{\ }_{j,x} F^{\ }_{j+\hat{y},x,\sigma} + \dots \ ,
\label{dimer}
\end{eqnarray}
where $\hat{x}$ ($\hat{y}$) denotes the unit lattice vector in x (y) direction and the dots indicate symmetry related and hermitian conjugate terms. 
Interestingly, these interaction terms are precisely equivalent to the bosonic and fermionic dimer resonance terms introduced in the generalized quantum dimer model of Ref.~\cite{Punk2015}. These terms are depicted graphically in Fig.~\ref{fig:dimer}. Here $\Delta^\dagger_{j,\eta} = (f^\dagger_{j \uparrow} f^\dagger_{j+\hat{\eta} \downarrow} - f^\dagger_{j \downarrow} f^\dagger_{j+\hat{\eta} \uparrow})/\sqrt{2}$ is the creation operator of a bosonic spin singlet dimer and $F^\dagger_{j,\eta,\sigma} =  (f^\dagger_{j \sigma} b^\dagger_{j+\hat{\eta}} + f^\dagger_{j+\hat{\eta} \sigma} b^\dagger_{j})/\sqrt{2}$ creates a fermionic dimer representing a spinon-holon bound state. The term $\sim a^\Delta_3$ in the first line then corresponds to the Rokhsar-Kivelson singlet resonance \cite{RokhsarKivelson}, whereas the other two are resonances between fermionic and bosonic dimers. Note that the dimer model has a hard-core constraint where each lattice site is part of precisely one dimer. In our case this hard-core dimer constraint directly follows from the particle number constraint in Eq.~\eqref{constraint} and the dynamics generated by the resonance terms in Eq.~\eqref{dimer} obey it. Another consequence of this constraint is that no quadratic hopping terms for $F_{ij\sigma}$ or $\Delta_{ij}$ are allowed, even though such terms would appear within a naive mean field decoupling of the $a^{F\chi}_{1,2}$ and $a^{\Delta \chi}_{1,2}$ terms in Eq.~\eqref{FLstaraction}. Lastly, we note that the same relation between the electron operator and the dimer operators shown in Eq.~\eqref{electron} was derived in Ref.~\cite{Punk2015} by computing matrix elements of the electron operator in the dimer Hilbert space.

\begin{center}
\begin{figure*}
\includegraphics[width= \textwidth]{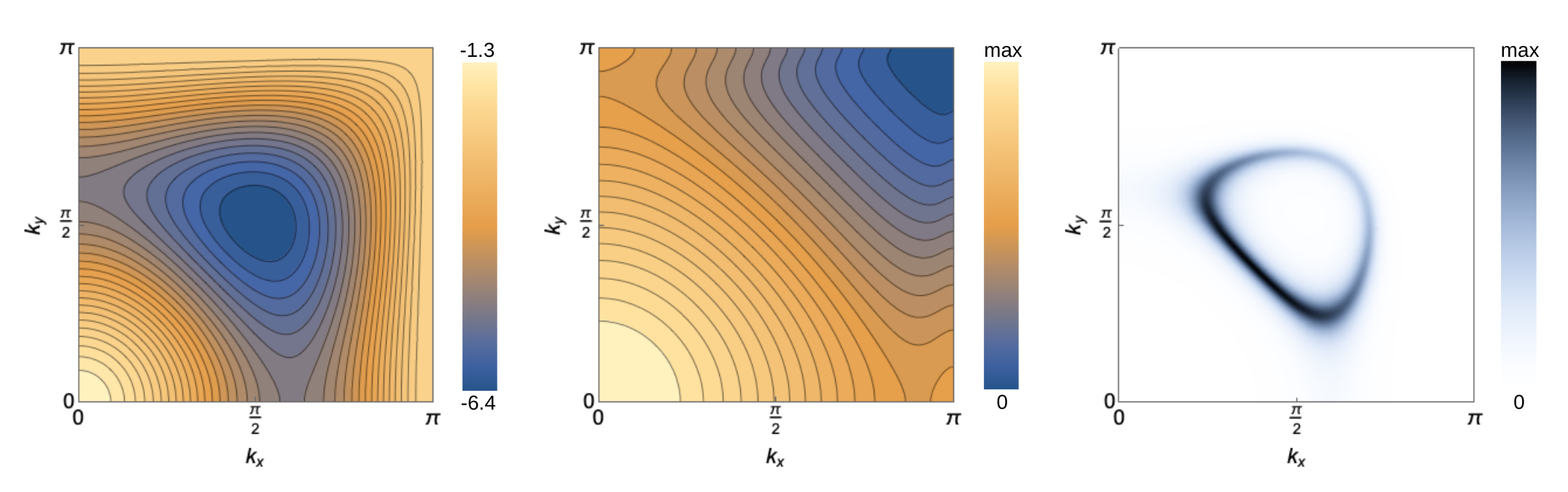}
\caption{Contour plots of the dispersion $E_-(\mathbf{k})$ (left) and electronic quasiparticle weight $Z_{\mathbf{k},-}$ (middle) in the $\mathbb{Z}_2$-FL* phase as function of momenta in one quadrant of the Brillouin zone. Right: density plot of the electron spectral function from Eq.~\eqref{elspecfunc} at the Fermi energy $\omega=0$ as function of momenta, with the delta function replaced by a Lorentzian with finite width. Parameters: $t_1=-1$, $t_2=2$, $t_3=-0.6$ and $\mu_F=-5.6$.}
\label{fig3}
\end{figure*}
\end{center}

A detailed numerical study of this dimer model in Refs.~\cite{Punk2015,Huber2018} as well as the exact analytic solution found in Ref.~\cite{Feldmeier2018} show that the symmetric ground state is indeed a $U(1)$-FL* with a sharp, small electronic Fermi surface. In a parameter regime relevant for the cuprates small Fermi pockets appear in the vicinity of momenta $\mathbf{k} = (\pm\pi/2, \pm\pi/2)$ with a Fermi volume proportional to the density of doped holes. Moreover, the electronic quasiparticle weight is finite and this theory exhibits well defined electronic quasiparticle excitations, even though Eq.~\eqref{electron} naively suggests that the electron spectral function is a convolution of the $\Delta$ and $F$ spectral functions, which should not exhibit a sharp quasiparticle peak. The reason is that the propagators of the $\Delta_{ij}$ and $F_{ij\sigma}$ fields remain local and do not acquire a dispersion due to the hard-core constraint. Interestingly the electronic quasiparticle weight is anisotropically distributed around the Fermi surface, giving rise to the appearance of Fermi arc-like features in the electron spectral function. We will find that the electron spectral function in the $\mathbb{Z}_2$-FL* phase discussed in the next section has very similar properties.

\subsection{$\mathbb{Z}_2$-FL* on the square lattice}

Here we study properties of the theory in Eq.~\eqref{FLstaraction} on the square lattice at the uniform RVB saddle point $\langle \chi_{j,\eta} \rangle = \chi$ and for $\langle \Delta_{j,\eta} \rangle \neq 0$. 
The action for the spinon pairing field $\Delta_{j, \eta}$ permits different non-trivial, translationally invariant saddle point solutions.
Here we focus on the extended $s$-wave case, where  $\langle \Delta_{j,x} \rangle = \langle \Delta_{j,y} \rangle = \Delta$. We comment on differences for a $d$-wave paired state with  $\langle \Delta_{j,x} \rangle = - \langle \Delta_{j,y} \rangle = \Delta$ at the end of this section.

For $\langle \chi_{j,\eta} \rangle = \chi$ and $\langle \Delta_{j,\eta} \rangle = \Delta$ the theory in Eq.~\eqref{FLstaraction} takes the form of a simple hopping Hamiltonian for the $F$ fermions. We keep three hopping terms within a tight-binding-like expansion (following the reasoning in Ref.~\cite{Punk2015}) and the effective mean-field Hamiltonian for the $\mathbb{Z}_2$-FL* is given by
\begin{eqnarray}
H_{\mathbb{Z}_2-\text{FL*}} &=&  - t_1 \sum_{j,\sigma} F^\dagger_{j+\hat{y},x,\sigma} F^{\ }_{j,x,\sigma} - t_2 \sum_{j,\sigma} F^\dagger_{j,y,\sigma} F^{\ }_{j,x,\sigma}  \notag \\
&& - t_3 \sum_{j,\sigma} F^\dagger_{j+\hat{y},y,\sigma} F^{\ }_{j,x,\sigma}  + \dots \ ,
\label{tightbinding}
\end{eqnarray}
where dots again denote hermitian conjugate, symmetry related, as well as possible longer range hopping terms. The corresponding hopping amplitudes are given by $t_1 = - a^{F\Delta}_2 |\Delta|^2 - a^{F\chi}_2 |\chi|^2$,  $t_2 = - a^{F\Delta}_1 |\Delta|^2 - a^{F\chi}_1 \chi^2$ and $t_3$ follows from a higher order loop term involving two elementary plaquettes.

\begin{center}
\begin{figure*}
\includegraphics[width= \textwidth]{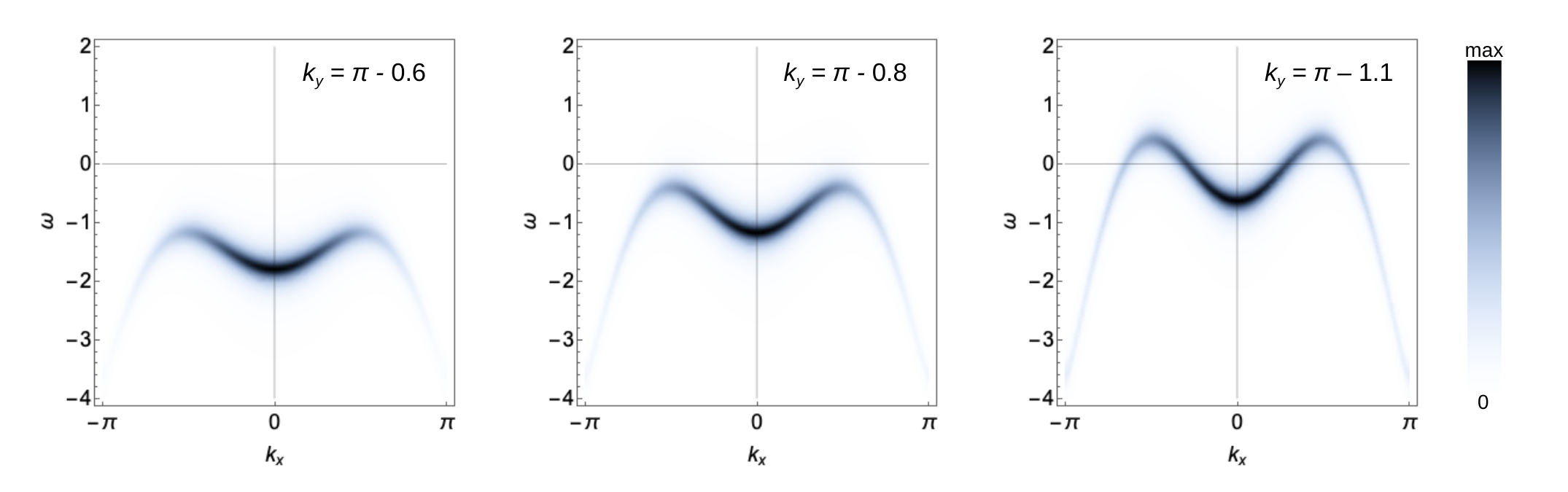}
\caption{Density plots of the electron spectral function $A_\text{electron}(\mathbf{k},\omega)$ in the $\mathbb{Z}_2$-FL* phase from Eq.~\eqref{elspecfunc} as function of momentum $k_x$ and frequency $\omega$, for three different values of $k_y$. The antinodal gap closes from below the Fermi energy ($\omega=0$) as $k_y$ decreases, leading to the formation of Fermi arc-like features close to $\mathbf{k} =(\pm \pi/2, \pm \pi/2)$. Parameters are the same as in Fig.~\ref{fig3} and the delta function in Eq.~\eqref{elspecfunc} was again replaced by a Lorentzian.}
\label{fig4}
\end{figure*}
\end{center} 

Even though the $F$ fermions are not gauge neutral, their Fermi surface is directly imprinted on the electronic Fermi surface. Indeed, for $\langle \Delta \rangle \neq 0$ we can use Eq.~\eqref{electron} to write the electron annihilation operator in momentum space as
\begin{equation}
c^{\ }_{\mathbf{k} \sigma} \sim \Delta  \sum_{ \eta=x,y} F^\dagger_{-\mathbf{k} \eta \bar{\sigma}} \big( 1+e^{i k_\eta }\big) \ .
\label{elZ2}
\end{equation}
It is straightforward to see that the single electron spectral function then takes the form
\begin{eqnarray}
A_\text{electron}(\mathbf{k}, \omega) &=& Z_\mathbf{k,-} \, \delta(\omega+E_-(-\mathbf{k})-\mu_F) \notag \\
&& +  Z_\mathbf{k,+} \, \delta(\omega+E_+(-\mathbf{k})-\mu_F) \ ,
\label{elspecfunc}
\end{eqnarray}
where $\mu_F$ is the chemical potential for $F$ fermions and 
\begin{equation}
E_\pm(\mathbf{k}) =  -t_1s_\mathbf{k} \pm \sqrt{t_1^2 d_\mathbf{k}^2 + 16 f_\mathbf{k}^2 \left[t_2 + 2 t_3 (s_\mathbf{k} - 1) \right]^2 } 
\label{dispersionE}
\end{equation}
denotes the tight-binding dispersions of the two bands obtained after diagonalising Eq.~\eqref{tightbinding}, where we defined $s_\mathbf{k} = \cos k_x + \cos k_y$,  $d_\mathbf{k} = \cos k_x - \cos k_y$ and $f_\mathbf{k} = \cos k_x/2  \, \cos k_y/2$. Note that there are two bands because the $F$ fermions reside on the links of the square lattice. The factors $Z_{\mathbf{k},\pm}$ determine the weight of the electronic quasiparticle peak and can be written as
\begin{equation}
Z_{\mathbf{k},\pm} \sim \sum_{\eta,\eta'} \left(1+e^{i k_\eta} \right) \left(1+e^{-i k_{\eta'}} \right) v^*_{\pm,\eta}(-\mathbf{k}) v_{\pm,\eta'}(-\mathbf{k}) \ ,
\label{eqZ}
\end{equation}
where the $v$'s are coefficients of the $2\times2$ matrix which diagonalises \eqref{tightbinding} in momentum space:
\begin{equation}
\begin{pmatrix}
F_{\mathbf{k},x,\sigma} \\
F_{\mathbf{k},y,\sigma} 
\end{pmatrix} =
\begin{pmatrix}
v_{+,x}(\mathbf{k}) & v_{-,x}(\mathbf{k})  \\
v_{+,y}(\mathbf{k}) & v_{-,y}(\mathbf{k})  \\
\end{pmatrix}
\begin{pmatrix}
F_{\mathbf{k},+,\sigma} \\
F_{\mathbf{k},-,\sigma} 
\end{pmatrix} \ .
\end{equation}
At low hole doping only the lower band will be occupied and in Fig.~\ref{fig3} we show contour plots of the corresponding dispersion $E_-(\mathbf{k})$ and quasiparticle weight $Z_{\mathbf{k},-}$. Here we've chosen values for the hopping matrix elements $t_1=-1$, $t_2=2$, $t_3=-0.6$, which have been estimated in Ref.~\cite{Punk2015}. Note that the dispersion has minima around $\mathbf{k} \simeq (\pm \pi/2, \pm \pi/2)$, leading to small hole-pockets centered around these momenta. Moreover, the quasiparticle weight is anisotropic and falls off towards the Brillouin zone corners, leading the appearance of Fermi arc-like structures in the electron spectral function, shown in Fig.~\ref{fig3}, where the outer side of the Fermi pocket has much lower spectral weight than the inner side. Note that this anisotropy is mainly due to the $e^{i k_\eta}$ factors in Eq.~\eqref{eqZ}, which are responsible for a vanishing of the quasiparticle weight at the Brillouin zone corners. These factors originate from the fact that the $F$ fermions live on the links of the lattice, see Eq.~\eqref{elZ2}.

In Fig.~\ref{fig4} we show cuts of the electron spectral function as function of momentum $k_x$ and frequency $\omega$ for three values of $k_y$ from the antinodal towards the nodal region. Note that the pseudo-gap at the antinodes closes from below the Fermi surface, in accordance with ARPES measurements \cite{He2011}.

Lastly we briefly mention differences for a $d$-wave paired $\mathbb{Z}_2$-FL* with $\langle \Delta_{j,x} \rangle = - \langle \Delta_{j,y} \rangle = \Delta$. In this case the $t_2$ and $t_3$ amplitudes in Eq.~\eqref{tightbinding} are modified. While the $t_1$ amplitude is identical for the two mean-field solutions for $\Delta$, the sign of the $t_2$ amplitude arising from the $a^{F\Delta}_1$ term in Eq.~\eqref{FLstaraction}, as well as the sign of the $t_3$ amplitude arising from the $a^{F\Delta}_2$ term is opposite. Note that the dispersion in Eq.~\eqref{dispersionE} is invariant under a combined sign change of $t_2$ and $t_3$.  Moreover, the expression for the electron operator in Eq.~\eqref{elZ2} has an opposite sign for the $\eta=y$ component and the quasiparticle weight is modified accordingly. Combined with a sign change of $t_2$ and $t_3$, the quasiparticle weight is identical to the extended $s$-wave paired case.

\section{Gauge fluctuations}
\label{sec:fluct}
 
In this section we study the model in Eq.~\eqref{FLstaraction} beyond the saddle point approximation for the bond field $\chi_{ij}$. We focus on important  
phase fluctuations of $\chi_{ij}$ restricted to nearest neighbor bonds for square lattice systems. For the following argument it is important to note that the fields $\chi_{ij}$ have a direction, i.e.~$\chi_{ji} = \bar{\chi}_{ij}$.
We can straightforwardly derive an effective theory for the $F$ fermions and the spinon pair field $\Delta$ coupled to a $U(1)$ gauge field by considering how phase fluctuations of $\chi_{ij}$ couple to the $F$ fermions and $\Delta$. For the uniform RVB phase on the square lattice we write $\chi_{j,\eta} = \chi \, e^{i A_{j}^{\eta}}$, where $A_j^\eta$ parametrizes the phase fluctuations of $\chi_{j,\eta}$ on the lattice bond emanating in $\eta=x,y$ direction from lattice site $j$.  The terms $\sim a^\chi_{3}$ in the first line of Eq.~\eqref{FLstaraction} then take the form
\begin{equation}
a^\chi_{3} \chi^4 \sum_j \exp i \left( A_{j}^{x}+A_{j+\hat{x}}^{y} - A_{j+\hat{y}}^{x}-A_{j}^{y} \right) + \text{h.c.} \ ,
\end{equation}
which corresponds to the elementary Wilson loop of a $U(1)$ lattic gauge theory. In the continuum limit we immediately obtain
\begin{equation}
2 a^\chi_{3} \chi^4  \int d^2x \, \cos(\partial_x A^y -\partial_y A^x) \ ,
\end{equation}
i.e.~the usual the Maxwell term for a compact $U(1)$ gauge theory in $2+1$ dimensions. The role of the time component $A^\tau_j$ of the gauge field is taken by the Lagrange multiplier terms $A^\tau_j \equiv \lambda_j$ in Eq.~\eqref{FLstaraction}.

The interaction between the $F$ fermions and $\Delta$ with the $U(1)$ gauge field follows from the terms $\sim a^{F\chi}_{1,2}, a^{\Delta\chi}_{1,2}$ in Eq.~\eqref{FLstaraction}: 
\begin{eqnarray}
&& a^{F\chi}_1 \chi^2 \sum_{i,\sigma} \bar{F}_{i,y,\sigma} F_{i,x,\sigma} \, e^{i (A^y_{i+x}-A^x_{i+y})}  + \dots \notag \\ 
&+& a^{\Delta \chi}_1 \chi^2 \sum_{i} \bar{\Delta}_{i,y} \Delta_{i,x} e^{i (A^y_{i+x}-A^x_{i+y})} + \dots \ ,
\end{eqnarray}
where the dots again denote symmetry related terms.
Carefully taking the continuum limit by expanding in gradients of the fermionic field and powers of the gauge field, 
the low energy theory of the FL* indeed describes charge-2 fermions $F$ as well as a charge-2 spinon pair field $\Delta$, which plays the role of a Higgs field, minimally coupled to a $U(1)$ gauge field. The Lagrangian density takes the form
\begin{eqnarray}
\mathcal{L}_\text{FL*} &=& \bar{F} \left[ (\partial_\tau - i 2 A^\tau) - (\nabla -i 2 \mathbf{A})^2 -\mu_F \right] F \notag \\
&& + \bar{\Delta} \left[ (\partial_\tau - i 2 A^\tau) - (\nabla -i 2 \mathbf{A})^2 + a^\Delta_1 \right] \Delta \notag \\
&& + a^\Delta_{2} |\Delta|^4 + S_\text{Maxwell}[A^\mu] \ ,
\label{gaugeFLs}
\end{eqnarray}
where $\mathbf{A} = (A^x,A^y)$ denotes the spatial components of the vector potential. For $\langle \Delta \rangle=0$ we thus obtain a $U(1)$-FL*, where a finite density of $F$ fermions is coupled to a $U(1)$ gauge field. Condensing the spinon pair field by setting $a^\Delta_1<0$ gaps out the $U(1)$ gauge field via the Higgs mechanism and the resulting phase is a $\mathbb{Z}_2$-FL*, where the $F$ fermions are coupled to an Ising gauge field. 

Due to the compactness of $A^\mu$ the $U(1)$ gauge theory allows for monopole excitations, which lead to confinement in the absence of matter fields. The presence of a Fermi surface, in our case of $F$ fermions, has been argued to suppress monopoles, however, and the theory in Eq.~\eqref{gaugeFLs} is expected to be deconfined \cite{Hermele2004}.

 \section{Quantum phase transitions}
 \label{sec:trans}

Quantum phase transitions between fractionalized Fermi liquids and an ordinary Fermi liquid (FL) or a superconductor (SC) are driven by the condensation of holons $b$, as indicated in the schematic phase diagram in Fig.~\ref{fig0}. For $\langle b \rangle = 0$ the electron Fermi surface coincides with the small $F$ Fermi surface of spinon-holon bound states and we are in an FL* phase, as discussed in Sec.~\ref{sec:fls}. By contrast, if the holons are condensed, $\langle b \rangle \neq 0$, we are either in an ordinary Fermi liquid phase, where the electron Fermi surface coincides with the large Fermi surface of spinons $f$, or the spinons are paired and the ground state is a superconductor. Here we propose theories for both transitions and briefly discuss their properties as well as potential problems. Similar quantum phase transitions in the context of Kondo-Heisenberg models have been discussed in Refs.~\cite{SenthilVojtaSachdev, SenthilVojtaSachdev2}.
 
\subsection{$\mathbb{Z}_2$-FL* to SC transition} 

The fermionic spinons don't play an important role at the $\mathbb{Z}_2$-FL* to superconductor transition. For the extended $s$-wave paired state the spinons remain fully gapped throughout the transition. Even though the $d$-wave paired state has gapless spinon excitations at four nodal points in the Brillouin zone, we don't expect them to play a prominent role at the transition. The important low energy degrees of freedom are the bosonic holons $b$, as well as the fermionic spinon-holon bound states $F$. Integrating out the spinons from the theory in Eq.~\eqref{lagrangian2} generates interaction terms between holons and the $F$ fermions of the form $\sim \bar{\Delta}_{ik} \, \bar{b}_j \bar{b}_\ell F_{ij \sigma} F_{k \ell \bar{\sigma}} + \text{h.c.}$ as well as $\sim \chi_{ik} \, \bar{F}_{k \ell \sigma} \bar{b}_j b_\ell F_{ij \sigma}$. Upon holon condensation the former interaction term induces pairing of the $F$ fermions, whereas the latter corresponds to a hopping term. This is expected from Eq.~\eqref{lagrangian2}, since the term in the last line hybridizes the $f$ and $F$ fermions in the presence of a holon condensate. 

Since the spinon pairing field $\Delta$ gaps out the $U(1)$ gauge field, as described in the previous section, the low energy action for the $\mathbb{Z}_2$-FL* to SC transition takes the form $S = S_b + S_F + S_\text{int}$ with
\begin{eqnarray}
S_b &=&  \int_{\tau,x}  \bar{b} \big(\partial_\tau - c \, \nabla^2 +s \big) b + u |b|^4  \ \ \ \ \ \\
S_F &=& \int_{\tau,x} \bar{F}_\sigma \big( \partial_\tau - \nabla^2 - \mu_F \big) F_\sigma  \\
S_\text{int} &=& \lambda \int_{\tau,x} \big( b^2 \bar{F}_\sigma \bar{F}_{\bar{\sigma}} +\text{h.c.} \big) \ .
\end{eqnarray}
The phase transition can be tuned via the boson mass term $s$. This theory has been discussed in detail in Ref.~\cite{Je1991}, where it was shown that the transition can be continuous, if the microscopic interaction between $F$ fermions is repulsive.

\subsection{$U(1)$-FL* to FL transition}

Here we discuss the quantum phase transition between the $U(1)$-FL* phase and an ordinary Fermi liquid and highlight important differences to Ref.~\cite{SenthilVojtaSachdev2} as well as potential problems with the low energy description of the $U(1)$-FL* phase.
For this transition the important low energy degrees of freedom are the spinons $f$, holons $b$, and their bound states $F$. These three degrees of freedom interact via the three-point interaction from the last line of Eq.~\eqref{lagrangian2}, as well as via the $U(1)$ gauge field described in Sec.~\ref{sec:fluct}. In the Fermi liquid phase, where $\langle b \rangle \neq 0$, the three-point interaction term hybridizes the $F$ and the $f$ electrons and the electron Fermi surface coincides with the large spinon Fermi surface. Note that these arguments are only relevant for the uniform RVB state, where a large spinon Fermi surface is indeed present. 

\begin{center}
\begin{figure}
\includegraphics[width=0.6 \columnwidth]{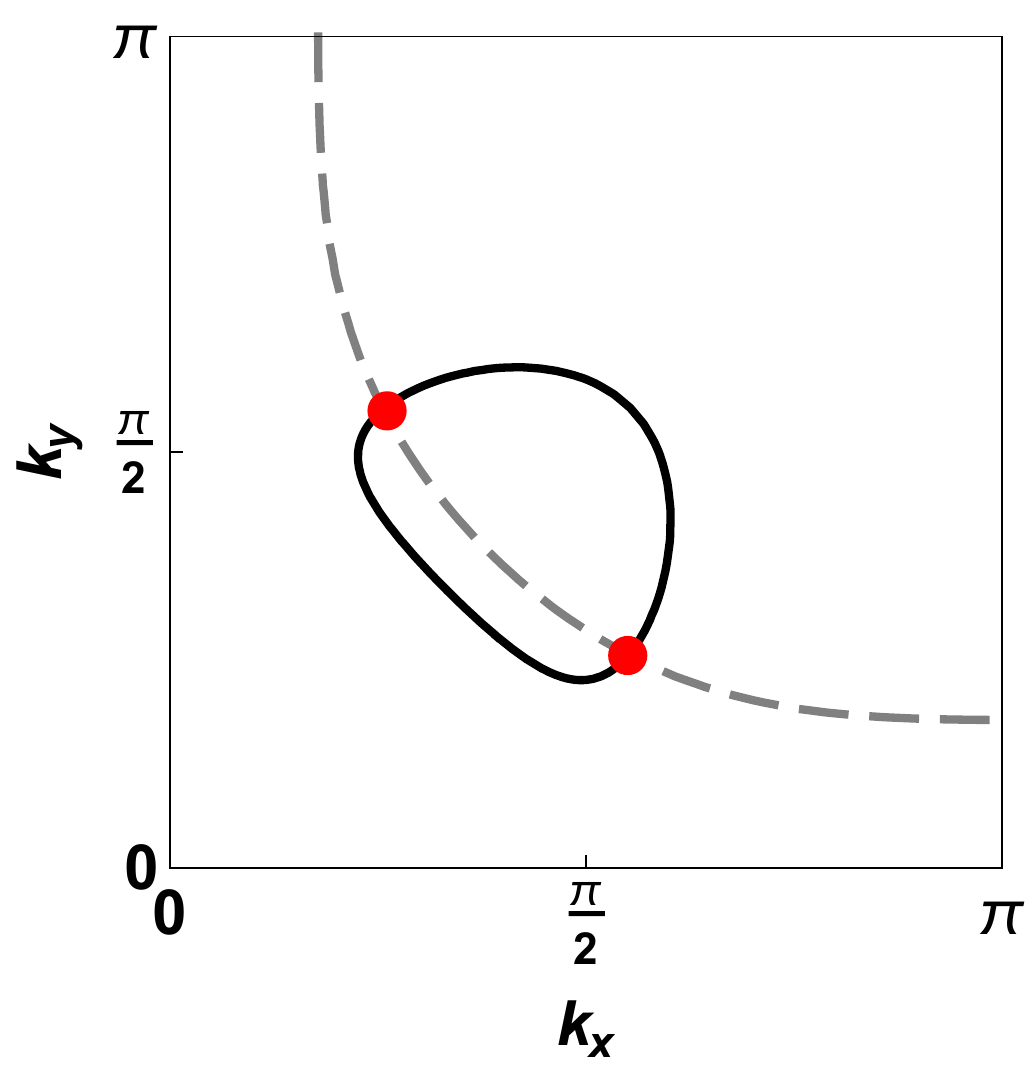}
\caption{Hot spots in the theory for the $U(1)$-FL* to FL quantum phase transition. The solid line indicates the small Fermi surface of fermionic spinon-holon bound states $F$, whereas the dashed line represents the large Fermi surface of spinons $f$. }
\label{fig:hotspot}
\end{figure}
\end{center}

Assuming holon condensation at momentum $\mathbf{k}=0$, the holons scatter off fermions via the three-point interaction with a small momentum transfer. The $f$ and $F$ fermions thus interact strongly at hot spots in momentum space, where their Fermi surfaces intersect. In contrast to Ref.~\cite{SenthilVojtaSachdev2} this appears to be a generic situation in the case of cuprates, where the small Fermi surface of $F$ fermions in the vicinity of $\mathbf{k} = (\pm \pi/2, \pm \pi/2)$ intersects with the large spinon Fermi surface, as indicated in Fig.~\ref{fig:hotspot}. A seemingly possible low energy action for a single hot-spot would take the form
$S = S_b + S_f + S_F + S_\text{int}$
with
\begin{eqnarray}
S_b &=& \int_{\tau,x}  \bar{b} \big(\partial_\tau-iA^\tau - c \, (\nabla-i \mathbf{A})^2 +s \big) b + u |b|^4  \ \ \ \ \ \\
S_f &=&  \int_{\tau,x} \bar{f}_\sigma \big(\partial_\tau-iA^\tau - i \, \mathbf{v}_f \cdot (\nabla-i \mathbf{A}) \big) f_\sigma \\
S_F &=& \int_{\tau,x} \bar{F}_\sigma \big( \partial_\tau-i2A^\tau -i \, \mathbf{v}_F \cdot (\nabla-i2 \mathbf{A}) \big) F_\sigma  \\
S_\text{int} &=& \lambda  \int_{\tau,x} ( \bar{F}_\sigma  f_\sigma b +  \bar{b} \bar{f}_\sigma F_\sigma ) \ .
\end{eqnarray}
Here we've chosen a local coordinate system where $\mathbf{v}_f$ and $\mathbf{v}_F$ denote the local Fermi velocities of $f$ and $F$ fermions at the hot spot. Even though this theory has a very similar structure as the one discussed in Ref.~\cite{SenthilVojtaSachdev2} for the $U(1)$-FL* to FL transition, the main difference is that our $F$ fermions, which take the role of the conduction electrons in Ref.~\cite{SenthilVojtaSachdev2}, are gauge charged. For this reason the dispersion term in the action $S_F$ for the $F$ fermions appears to be highly problematic: as pointed out in Sec.~\ref{sec:u1fls}, the $F$ fermions do not acquire a dispersion in the lattice scale description of the $U(1)$-FL* due to the hard-core dimer constraint. Indeed, $F$ fermions with a dispersion cannot give rise to a FL* phase, as Eq.~\eqref{electron} would preclude the existence of a sharp electronic Fermi surface. The continuum limit taken in the low energy theory above thus does not properly describe the $U(1)$-FL* phase. 

Finding a valid low energy theory for the $U(1)$-FL* to FL transition based on our $U(1)$ slave boson approach thus remains an open problem.
Interestingly, a possible theory for this transition has been developed recently in Ref.~\cite{Zhang2020}, using a different parton construction. It would be interesting to see if a connection can be established between the two approaches, potentially via a generalized $SU(2)$ slave boson construction \cite{Wen1996}.

\section{Discussion and Conclusions}

We modified the standard $U(1)$ slave boson theory of the $t$-$J$ model to account for spinon-holon bound states. This theory has fractionalized Fermi liquid ground states which capture essential features of the nodal-antinodal dichotomy in the metallic pseudogap phase of underdoped cuprates. In particular, the pseudogap in the antinodal region of the Brillouin zone at momenta $\mathbf{k}=(0,\pm \pi)$ and $(\pm \pi,0)$ closes from below the Fermi surface when moving towards the nodal region close to momenta $\mathbf{k} = (\pm \pi/2,\pm \pi/2)$, in accordance with experimental observations. Moreover, the anisotropic electronic quasiparticle weight around the Fermi pockets in the vicinity of the nodal points makes these pockets appear as Fermi arcs in photoemission experiments. 

Our theory can be used as starting point to investigate possible symmetry broken phases and to compute further observables that can be compared to experimental data. Generalizations to the $SU(2)$ slave boson construction of Wen and Lee \cite{Wen1996} are possible as well and might provide interesting connections to recent work on $SU(2)$ gauge theories for cuprates \cite{Scheurer2018, Sachdev2019}.

\acknowledgements
We acknowledge helpful discussions with Erez Berg, Sebastian Huber, Inti Sodemann and in particular with Subir Sachdev. This work is supported by the Deutsche Forschungsgemeinschaft (DFG, German Research Foundation) via the Munich Center for Quantum Science and Technology (MC-QST) - EXC-2111-390814868.   
 
%

\bibliography{refs}
\bibliographystyle{apsrev4-1}

\end{document}